\documentclass[11pt,twoside]{article}

\usepackage{asp2006}
\usepackage{epsf}
\usepackage{psfig}
\usepackage{lscape}

\begin{document}
\title{ Tidal Tales of Minor Mergers: Star Formation in the Tidal Debris of
 Minor Mergers}   
\author{Karen A. Knierman}   
\affil{School of Earth and Space Exploration, Arizona State University, Tempe, AZ 85287}    

\begin{abstract} 
How does the tidal debris of minor galaxy mergers contribute to
structures in spiral galaxies or in the intergalactic medium? While
major mergers are known to create structures such as tidal dwarf
galaxies and star clusters within their tidal debris, less is known
about minor mergers (mass ratios between a dwarf galaxy and disk
galaxy of less than one-third) and their tidal debris. This work
surveys 6 nearby minor mergers using optical broad-band and $H\alpha$
narrow-band imaging to characterize star formation in their tidal
debris. Young star clusters with ages less than the dynamical age of the tidal tails are found in all 6 mergers, indicating that the star clusters formed in situ.  
Even if minor mergers contribute less tidal debris per
interaction than major mergers, they are more common and possibly
contribute structure to all types of galaxies and to the intergalactic
medium throughout the history of the universe.

\end{abstract}



\section{Introduction}
Does the tidal debris of minor mergers contribute structures to halos of spiral galaxies or IGM?  Though major mergers are known to create structures from star clusters to tidal dwarf galaxies in their debris \citep{duc,knierman}, less is known about minor mergers and their tidal debris.  Yet, minor mergers are more common and were presumably important in the early universe.  Structures in minor mergers' tidal debris may be captured by the more massive galaxy and create structures in the galaxy's halo like those found in the Milky Way \citep{ibata}, or contribute to the pollution of the IGM.  Tidal debris can also give insight into unusual environments of star formation and how gas behaves in galaxy interactions. Studying the formation of structure in the tidal debris of minor mergers can help us understand a common process in galaxy formation, particularly disk and halo formation.

\section{Background}
Compared to major mergers, less is known about minor mergers (mass ratios between a dwarf galaxy and a disk galaxy of  less than 0.3) and their tidal debris.  Previous work includes VLT observations of NGC 6872 \citep{bastian} and HST observations of UGC 10214 \citep{tran, deGrijs}.  These studies found the tidal debris to contain blue star clusters ($U-B \sim -0.3$; Bastian et al. 2005 and $B-V \sim -0.2$; de Grijs et al. 2003) with young ages indicating they were formed within the tidal tail(s).

\section{Methods}
Using the 1.8 m VATT, deep optical images of 15 minor mergers were taken in $UBVR$ and narrow-band $H\alpha$ and reduced using standard IRAF tasks.  Point sources were found using DAOFIND.  Final star cluster candidates were selected using the 3DEF code \citep{bik} and Bruzual \& Charlot (2003) SSP models with a metallicity of $0.4Z_{\odot}$.  These SED fits gave an age, mass, and extinction for each star cluster candidate.

\section{Results}
Results from the closest six out of the full sample of 15 minor mergers show that star clusters in the tidal debris have ages ranging from a few Myr to greater than a Gyr  as well as masses ranging $10^{3.5} - 10^8 M_\odot$ (Figure 1).  There may be some contamination from foreground stars and background galaxies.  Table 1 shows properties of the minor mergers such as merger stage, as well as the dynamical age of the tail, number of star cluster candidates (SCC), mean or median values of $M_B$, age, mass, and extinction, and the total mass of the star cluster candidates in the tidal debris area.  

In each merger, there are SCC with ages less than the tail age, which indicates that the star cluster formed in the tidal debris and was not pulled out from the central regions of the galaxies.  
Comparing the merger stages with age, early stage mergers have slightly younger median ages than late stage or merged systems, with the exception of UGC 260. The median age of clusters in this system are likely skewed by the presence of substantial $H\alpha$ flux contaminating the clusters. In general, the median cluster age appears to be 10-20\% of the dynamical age of the merger.  The lowest median mass of the star clusters are found in the early stage mergers of Arp 269 and Arp 279.  These two galaxies appear to be missing higher mass star clusters. This could indicate either that they have not yet formed more massive star clusters or that there is not enough material in their tails to form them.
 
 \begin{figure}[!ht]

\plotone{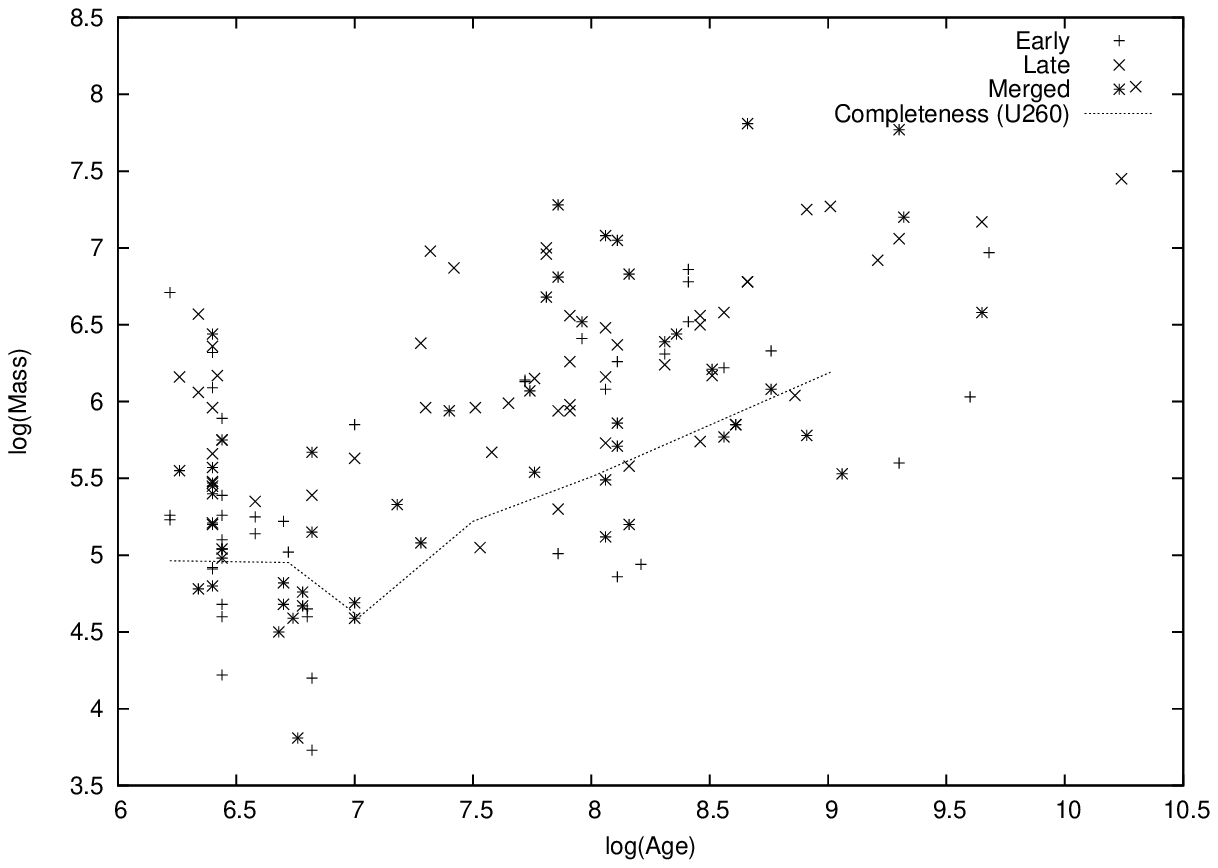}

\caption{Mass and Age for all star cluster candidates, sorted by merger stage, fit to \citet{bc03} SSP models by the 3DEF code \citep{bik}. The dotted line shows the completeness limit for UGC 260.}

\end{figure}

\begin{table}
\caption{Properties of Tidal Debris and Star Cluster Candidates}
\small
\begin{tabular}{cccccccc}
\tableline
\tableline
\noalign{\smallskip}
Debris & Merger & Tail Age & $\#_{SCC}$ & $M_B$\tablenotemark{b}  & Mass\tablenotemark{c}  & Age\tablenotemark{c}    & Total Mass\\
Area & Stage\tablenotemark{a} & Myr & & mag & $log[M_\odot]$ &  $log(yr)$  & $log[M_\odot]$ \\
\noalign{\smallskip}
\tableline
Arp269E & 1 & 72.8 & 12 & -7.89 & 4.99 & 6.82  & 7.07 \\
Arp269B &  & & 10 & -8.47 & 5.13 & 6.44 &  6.3 \\
Arp279W & 1 & 34.7 & 4 & -10.37 & 5.68 & 6.71 & 6.8 \\
Arp279B &  & 54.4 & 6 & -10.08 & 5.57 & 6.51 & 6.4 \\
UGC260N & 1 & 30.7 & 5 & -12.20 & 6.26 & 8.06  & 7.1 \\
UGC260S &  & 47.2 & 6  & -12.02 & 6.37 & 8.41  &7.3 \\
NGC2782E & 2 & 200.0 & 32 & -10.44 & 6.38 & 8.00  & 8.4 \\
NGC2782W &  & 320.9 & 19 & -9.86 & 5.75 & 7.32 & 7.6 \\
NGC3310 & 3 & 77.9 & 43 & -9.96 & 5.47 & 7.18  & 8.1 \\
NGC7479 & 3 &  118.5 & 10 & -12.13 & 6.42 & 7.91 & 8.0 \\

\tableline
\tableline
\end{tabular}
\tablenotetext{a}{1=early, 2=late, 3=merged}
\tablenotetext{b}{Mean value for debris area}
\tablenotetext{c}{Median value for debris area}

\end{table}

\subsection{Star Formation Rates in Tidal Debris}

\begin{table}
\caption{Properties of HII regions in the Tidal Debris of Minor Mergers}
\small
\begin{tabular}{cccccc}
\tableline
\noalign{\smallskip}
Debris & $L_{H\alpha,lim}$ & $N_{HII}$ & $L_{H\alpha,med}$ & $L_{H\alpha,tot}$ & $SFR(H\alpha)$\tablenotemark{a}  \\
 & $10^{37} erg s^{-1}$ & & $10^{37} erg s^{-1}$ & $10^{37} erg s^{-1}$ & $M_{\odot} yr^{-1}$\\\noalign{\smallskip}
\tableline
A269Br & 0.562 & 4 & 3.16 & 13.6 & 0.00108\\
A269Dw & 0.562 & 11 & 6.30 & 318 & 0.0248\\
A279Br & 0.0942 & 10 & 11.5 & 141 & 0.0112\\
U260N & 0.117 & 8 & 58.5 & 648 & 0.0512\\
U260S & 0.117 & 5 & 92.7 & 713 & 0.0563\\
N2782E & 0.568 & 5 & 27.4 & 214 & 0.0169\\
N2782W & 2.67 & 1 & 137 & 137 & 0.0108\\
N3310N & 0.0894 & 31 & 60.5 & 5520 & 0.436\\
N3310S & 0.0894 & 18 & 69.1 & 1570 & 0.124\\

N7479N\tablenotemark{c} & 4.47 & 15 & $>10$ & $>389$ & $> 0.0307$\\
\tableline
\tableline
N4676\tablenotemark{e} &  & & & 1100 & 0.087\\
N4038S\tablenotemark{e} & & & & 170 & 0.013\\
N7714/5Br\tablenotemark{e} & & & & 180 & 0.014\\
\tableline
\tableline
\end{tabular}

\tablenotetext{a}{Kennicutt 1998}
\tablenotetext{d}{Using HII regions from Jarrett et al. 2006, median, total, and SFR are taken from the more massive HII region.}
\tablenotetext{e}{Major Merger tidal debris from Smith et al. 1999 and references therein.}

\end{table}

In the sample of 10 debris regions in 6 minor mergers, we find that 8 debris regions host HII regions (see Table 2).  The debris regions of NGC 3310 have the highest median $H\alpha$ luminosity, total $H\alpha$ luminosity, and total star formation rate.  The tidal tails of UGC 260 have the next highest values behind NGC 3310. In comparing the minor merger $H\alpha$ luminosities, we find that they are of similar orders of magnitude to major mergers.  So star formation occurs at similar levels in both major and minor mergers.  In comparing merger stages, we find that early stage mergers in the sample are forming smaller HII regions than those in late or merged systems.

There is a weak trend in $H\alpha$ luminosity and thus star formation rate with merger stage, merged systems having a higher rate. However, UGC 260 has the second highest rate in the sample. This indicates that individual galaxy parameters are more important than merger stage.

\section{Conclusions}

These observations show that star clusters with ages less than the tail age are present in all sample tidal debris.  This suggests that star formation in minor merger tidal debris is ubiquitous.  Ages appear to correlate with dynamical age of the merger.   There is a weak trend in $H\alpha$ luminosity and thus star formation rate with merger stage, merged systems having a higher rate.  There may be delays of  $>10$ Myr in star formation episodes as two debris regions in the early stage systems did not harbor any HII regions.

\acknowledgements 
Thank you to the Galaxy Wars committee for providing conference funding and to the Arizona/NASA Space Grant Consortium.


\end{document}